\documentclass[fleqn,usenatbib]{mnras}

\usepackage{newtxtext,newtxmath}

\usepackage[T1]{fontenc}
\usepackage{ae,aecompl}

\usepackage{natbib}
\usepackage{xcolor}
\usepackage{etoolbox}
\usepackage{hyperref}
\usepackage{afterpage}
\usepackage{float}
\usepackage{multirow}
\usepackage{deluxetable}
\usepackage{soul}

\makeatletter
\def\@to{to}
\makeatother

\usepackage{graphicx}	
\usepackage{amsmath}	
\usepackage{amssymb}	

\newcommand{\siii}{\mathrm{Si}\:\textsc{ii}}
\newcommand{\siiv}{\mathrm{Si}\:\textsc{iv}}
\newcommand{\cii}{\mathrm{C}\:\textsc{ii}}
\newcommand{\civ}{\mathrm{C}\:\textsc{iv}}
\newcommand{\oi}{\mathrm{O}\:\textsc{i}}
\newcommand{\ewlim}{\mathrm{EW}\geq 50\:\:{\mbox{m\normalfont\AA}}}
\newcommand{\mA}{\mbox{m\normalfont\AA}}
\newcommand{\hmpc}{h^{-1}\mathrm{Mpc}}
\newcommand{\kms}{\:\:\mathrm{km\,s}^{-1}}

\newcommand{\msun}{M_\odot}

\title[Aligned Metal Absorbers]{Aligned Metal Absorbers and the Ultraviolet Background at the End of Reionization}

\author[C. Doughty et al.]{
Caitlin Doughty,$^{1}$\thanks{E-mail: cdoughty@nmsu.edu}
Kristian Finlator,$^{1}$
Benjamin D. Oppenheimer,$^{2}$
Romeel Dav\'e,$^{3,4}$
\newauthor
Erik Zackrisson$^{5}$
\\
$^{1}$Department of Astronomy, New Mexico State University, Las Cruces, NM 88001\\
$^{2}$CASA, Department of Astrophysical and Planetary Sciences, University of Colorado, 389-UCB, Boulder, CO 80309, USA\\
$^{3}$Institute for Astronomy, Royal Observatory, Edinburgh EH9 3HJ, UK\\
$^{4}$Department of Physics and Astronomy, University of the Western Cape, Bellville, Cape Town 7535, South Africa\\
$^{5}$Department of Physics and Astronomy, Uppsala University, 751 20 Uppsala, Sweden
}

\date{Accepted XXX. Received YYY; in original form ZZZ}

\pubyear{2017}

\begin{document}
\label{firstpage}
\pagerange{\pageref{firstpage}--\pageref{lastpage}}
\maketitle

\begin{abstract}
We use observations of spatially-aligned $\cii$, $\civ$, $\siii$, $\siiv$, and $\oi$ absorbers to probe the slope and intensity of the ultraviolet background (UVB) at $z \sim 6$.  We accomplish this by comparing observations with predictions from a cosmological hydrodynamic simulation using three trial UVBs applied in post-processing: a spectrally soft, fluctuating UVB calculated using multi-frequency radiative transfer; a soft, spatially-uniform UVB; and a hard, spatially-uniform ``quasars-only'' model.  When considering our paired high-ionization absorbers ($\civ$/$\siiv$), the observed statistics strongly prefer the hard, spatially-uniform UVB.  This echoes recent findings that cosmological simulations generically underproduce strong $\civ$ absorbers at $z>5$.  A single low/high ionization pair ($\siii$/$\siiv$), by contrast, shows a preference for the HM12 UVB, while two more ($\cii$/$\civ$ and $\oi$/$\civ$) show no preference for any of the three UVBs.  Despite this, future observations of specific absorbers, particularly $\siiv$/$\civ$, with next-generation telescopes probing to lower column densities should yield tighter constraints on the UVB.
\end{abstract}




\section{Introduction}
Hydrogen reionization is believed to have occurred during the redshift range $6<z<13$~\citep{beck15,planck16}, but the exact sources of the ionizing light remain uncertain.  Direct observations of bright quasars indicate that they are a primary source of ionizing photons at lower redshifts, but their number density decreases significantly at $z>4$~\citep{beck15,jia16,qin17,mant17}, implying that other ionizing sources become significant.  Recent observations suggest that faint quasars may have made an unexpectedly large contribution to the high-redshift UVB~\citep{gia15}, although these observations are not irreconcilable with galaxy-dominated reionization~\citep{jia16,ric16}.  However, the former conclusion is only true if the fraction of ionizing photons that escape $f_{esc}$ from quasars is unity, and this conventional assumption may not be correct~\citep{mic17,cri16}.  Further, other work has shown that reionization by faint active galactic nuclei would result in predictions of optical depth, neutral fractions, and ionizing emissivity that conflict considerably with observations~\citep{hass17}.  Meanwhile, galaxies may well have been abundant enough to drive reionization, but only if the faintest ones ($M_\mathrm{UV} > -15$) were abundant~\citep{fink12,liv17}.

One approach to identifying the ionizing sources leverages their relationship to the rapidly-growing UVB, which can be observed indirectly via its influence on the Lyman-$\alpha$ forest and on metal absorption systems.  While modeling the Lyman-$\alpha$ forest has long been an important way of studying the epoch of reionization and its aftermath~\citep{fan06,beck15}, the constraining power of metal absorbers has been neglected until relatively recently.  Broadly, the Lyman-$\alpha$ forest probes gas that is at or slightly below the mean density (with higher-order transitions probing higher densities~\citep{fur09}), while metal absorbers probe the overdense gas that is closely associated with galaxies~\citep{opp09,kea16}.  The sensitivity of metal absorbers to the UVB's spectral slope and their close spatial association with galaxies renders them a particularly promising probe of the ionizing population.  For example, a tendency for metals to be in highly-ionized phases such as $\civ$ indicates either a hard or an intense UVB.  Moreover, if the increased intensity exceeds what is allowed by analysis of the Lyman-$\alpha$ forest, then small-scale fluctuations in intensity or spectral hardness are implied.  Such strong fluctuations would naturally point toward a significant contribution from nearby galaxies, thereby constraining their contribution to their overall ionizing photon budget.  By contrast, if the observed metal absorber abundance ratios and the Lyman-$\alpha$ forest can be accounted for simultaneously in a model that assumes negligible contribution from galaxies, then a quasar-only reionization model is allowed.  Since Lyman-$\alpha$ is saturated for $z > 6$, the qualities of the UVB as measured from metal absorbers will be the most illuminating with regards to the ionizing sources.

The challenge in using metal absorbers to probe the UVB is that the metallicity, temperature, and UVB are all spatially-inhomogeneous.  These fluctuations are best accounted for using modeled absorption spectra from cosmological simulations.~\citet{opp09} implemented this idea in a simple post-processed fashion, showing that the statistics of velocity-aligned metal absorbers could distinguish between several trial UVBs.  In a recent work~\citep{fin16}, we explored the predicted abundance ratios of $\siiv$, $\civ$, and $\cii$ in a cosmological radiation hydrodynamic simulation.  This work improved on~\citet{opp09} by accounting for self-shielding in dense gas and treating small-scale UVB fluctuations self-consistently.  By varying the UVB in post-processing, it was argued that observations are inconsistent with a quasars-only UVB owing to its tendency to overproduce $\civ$.  Observations were also found to be in mild tension with the spatially-uniform~\citet{hm12} model (HM12), which \emph{under}produced $\civ$.  By contrast, a realistic, galaxy-dominated, spatially-fluctuating UVB whose amplitude was tuned to match observations of the Lyman-$\alpha$ forest was argued to reproduce observations well.  In this work, we build on that initial study in three ways.   First, we additionally consider $\oi$ and $\siii$ absorption.  Next, we relax the requirement that more than two ions be constrained simultaneously. Doing so results in a larger data sample to compare to simulations.  Finally, we focus our analysis on the overall aligned absorber fraction rather than on the distribution of abundance ratios.  Here, aligned absorbers are absorption lines of two different transitions lying less than 50$\kms$ apart along a QSO sightline.

In Section~\ref{sec:simulations}, we begin by describing the simulations, and the method used to define absorption systems.  The statistical analyses applied to high-ionization and low-ionization ions are introduced in Section~\ref{sec:analysis}.  We present the results in Section~\ref{sec:results} and discuss their implications in Section~\ref{sec:discussion}.  In Section~\ref{sec:predictions} we offer predictions regarding how changes in observational sensitivity may assist in distinguishing between these model UVBs in the future, as well as which transition pairs would be most beneficial to observe.  Finally, we present the overall conclusions of this venture in Section~\ref{sec:conclusions}.  All figures and statistics in this paper are derived from models and data at a redshift $z\sim6$.

\section{Simulations and Identification of Absorption Systems} \label{sec:simulations}
Our study builds on the preliminary discussion of metal absorbers in~\citet{fin16} and is based on a broadly similar simulation of early galaxy formation.  The simulation treats hydrodynamics, gravity, star formation, and feedback by discretizing the matter in a periodic box of length $12\:\:\hmpc$ into $2\times256^3$ particles and modeling their evolution from $z=199\rightarrow4$ using a custom version of {\sc Gadget-3}.  Our adopted cosmology is one in which $\Omega_M=0.3089$, $\Omega_\Lambda=0.6911$, $\Omega_b = 0.0486$, $h=0.6774$, $\sigma_8 = 0.8159$, and the index of the primordial power spectrum $n=0.9667$. 

Our simulation differs from that considered in~\citet{fin16} via an array of small improvements which we will present in an upcoming work (Finlator et al.\ 2018, in prep). We assume a~\citet{krou01} initial mass function.  We model metal enrichment from core collapse supernovae (SNe) using the~\citet{nomo06} yields assuming a 50\% hypernova fraction.  The latter assumption boosts the oxygen yield by $\sim20\%$ for low-metallicity SNe without significantly affecting C or Si yields.  For galactic outflows, we adopt a relation between mass-loading factor and stellar mass that emerges self-consistently from high-resolution simulations~\citep{mura15}.  Note that, in contrast to our previous work, the newly-adopted outflow scalings do not involve an empirical calibration.  It has previously been shown that, when folded into cosmological simulations, these scalings naturally lead to excellent agreement with the observed history of star formation~\citep{dave16}; we find the same result but do not reproduce it here.  The new outflow scalings suppress the star formation rate density at $z\sim6$ by roughly a factor of 2.5 with respect to the \textit{vzw} model adopted in~\citet{fin16}.

The way in which our version of {\sc Gadget-3} departs most significantly from the industry standard is in its UVB treatment: Rather than adopting a precomputed, spatially-uniform UVB, we decompose the UVB into galaxy and quasar contributions and model both using an on-the-fly continuum radiation transport solver~\citep{fin15,fin16}.  The galaxy contribution to the UVB is discretized into $32^3$ voxels and 24 independent frequency bins spaced evenly between 1--10 Ryd and is evolved using the local emissivity and opacity.  Each voxel's emissivity is computed from the local metallicity-weighted star formation rate density using the {\sc Yggdrasil} spectral synthesis code~\citep{zac11} with an imposed model for the ionizing escape fraction.  The latter depends only on redshift and is tuned to yield simultaneous agreement with observations from the Lyman-$\alpha$ forest and the observed optical depth to Thomson scattering (see below; note that this is the \emph{only} empirical calibration in our new simulations).  The local opacity owes to bound-free transitions in hydrogen and helium, whose ionization states are tracked using a nonequilibrium solver.

The QSO contribution to the UVB is likewise modeled on-the-fly using 24 frequency bins, but it is spatially-homogeneous and evolved using the volume-averaged opacity in each frequency bin.  We adopt the~\citet{luss15} continuum slope for the QSO emissivity, the~\citet{mant17} QSO redshift-dependent emissivity at 1 Ryd, and assume a 70\% ionizing escape fraction from QSOs~\citep{cri16}. We set the quasar emissivity to zero for $z\geq8$.  Note that current models predict an overall weak QSO contribution at $z=6$: In the~\citet{hm12} UVB, QSOs contribute 4.5\% of the total emissivity at 1 Ryd.  In our model, they likewise contribute 6\% of the total.

A frequency-dependent, subgrid self-shielding model attenuates both the galaxy and the QSO UVBs in the vicinity of dense gas and reduces the contribution of that gas to the opacity field.  An iterative solver assures that the UVB and the ionization field are consistent at each timestep.  In this way, our simulated UVB regulates the density and temperature of diffuse gas and drives a realistic reionization history: The predicted optical depth to Thomson scattering is 0.062, in good agreement with the observed range ($0.055\pm0.009$;~\citealt{planck16}).  Additionally, the UVB's amplitude is compatible with observations from the Lyman-$\alpha$ forest at $z\sim6$: The volume-averaged neutral hydrogen photoionization rate per second $\Gamma$ at $z=6.016$ is observed to be $\log_{10}(\Gamma)=-12.94\pm0.55$ (\citealt{cal11}; see also~\citealt{wyi11}).  Meanwhile, our simulation yields $\log_{10}(\Gamma)=-12.72$ and the HM12 UVB yields 
-12.60 at the same redshift.  Hence while both models are clearly in excellent agreement with the measurement, they are in even better agreement with one another.  We therefore adopt our simulated UVB without imposing any extra calibration.

In short, our new simulations offer improved agreement with the history of star formation and hydrogen reionization for the price of exactly one empirically-calibrated parameter, namely the ionizing escape fraction from galaxies.  Hence they are an excellent starting point for interpreting observations of reionization-epoch metal absorbers.

As in~\citet{fin16}, the goal of our inquiry is to ask to what extent observations of metal absorbers can be used to constrain the UVB.  To address this question, we cast sightlines through our simulation volume and compute, at the position of each 2 $\kms$ pixel, the local opacity owing to $\oi$, $\siii$, $\siiv$, $\cii$, and $\civ$ assuming ionization equilibrium with each of three trial UVBs as described in~\citet{fin15}.  The first trial UVB is the simulated UVB taken without modification.\footnote{This is an improvement with respect to~\citealt{fin16}, where it was necessary to adjust the UVB amplitude in post-processing in order to match HM12 at the Lyman limit.  In our updated simulations, this adjustment is unnecessary.}  The second is the~\citet{hm12} UVB.  The third is a model QSOs-only UVB, which we produce by neglecting the galaxy UVB and adjusting the QSO portion of the simulated UVB until its amplitude matches HM12 at the Lyman limit.  The simulated UVB contains spatial fluctuations~\citep[Figure 4 of][]{fin15} and a very weak QSO contribution; the HM12 model represents a similar mix of galaxy and QSO photons but is spatially-homogeneous; and the QSOs-only model is homogeneous on large scales (although when adopting it we include self-shielding), and it is much harder spectrally.  Note that the gas temperature and hydrogen ionization state are adopted without modification from the simulation.  This leads to slight inconsistencies when we adjust the UVB in post-processing (for example, a harder UVB should be associated with a hotter IGM).  However, we do not expect this to dominate our results because the high-ionization ions that we consider are predominantly photoionized, as has also been found in complementary studies~\citep{bir16}.  To mimic realistic spectra, we smooth the resulting spectra with a gaussian response function with full width at half-maximum of 10$\kms$ and add gaussian noise corresponding to a signal-to-noise of 50 per pixel.

We identify absorption lines in our synthetic quasar spectra as regions where the normalized flux drops $5\sigma$ (corresponding to a flux decrement of 10\%) below the continuum in at least three contiguous pixels.  Optical depth is then computed following~\citet{sava91}. We have verified that this fast line-identification approach yields results that are nearly indistinguishable from what we obtain using rigorous line-fitting~\citet{dav97}.  Following convention in the literature, we circumvent ambiguity in the decomposition of lines that are closely-located in velocity space by merging lines that are within 50 $\kms$ of one another into a single ``system''~\citep{song01}.  From the output, lines of the detection transition are sorted in order of descending strength.  Beginning with the strongest detected line, weaker lines are merged into the stronger lines if they fall within 50$\kms$ of one another, yielding a catalog of merged systems.  This process is repeated with all of the transitions that we consider.  Next, we identify one transition as the ``search'' transition and another as the ``aligned'' transition.  We consider only transition pairs that correspond to different ions.  Aligned systems are matched to search systems if they fall within 50$\kms$ of one another.  This method yields a catalog of matched systems that can be compared directly with observations.

\section{Analysis} \label{sec:analysis}
Our goal is to determine the compatibility of simulated and observed catalogs of matched absorbers.  The statistic that we focus on is the ``aligned absorber fraction'' $f$: Given a matched transition pair (as described in Section~\ref{sec:simulations}), this is the fraction of systems that are identified in the search transition, that are also detected in the aligned transition.
We calculate the aligned fraction in the data and in our three model catalogs (one for each trial UVB), and then use Poisson statistics to ask which UVB model the observed $f$ prefers.  While this is conceptually simple, complications arise owing to sightline-to-sightline variations in sensitivity, and to the presence of systems for which only upper limits are reported.  In this section, we discuss how we statistically compare the simulated and observed catalogs.

We begin by discussing how we evaluate the compatibility between observed and simulated absorption systems in the general case where $f>0$; i.e., the aligned transition is detected in at least one case.  Suppose, for concreteness, that we are studying the aligned absorber fraction for the $\civ$/$\siiv$ pair.  We define the the conditional $\siiv$ column density distribution (CCDD) as the probability density $dP/dN_{\siiv}(N_\mathrm{\siiv} \mid N_{\mathrm{\civ, min}})$ that a system's $\siiv$ column density lies between $N_{\mathrm{\siiv}}$ and $N_{\mathrm{\siiv}} + dN_{\mathrm{\siiv}}$ when considering only systems whose $\civ$ column density exceeds some threshold $N_\mathrm{\civ, min}$.
\begin{figure}
\includegraphics[width=0.5 \textwidth]{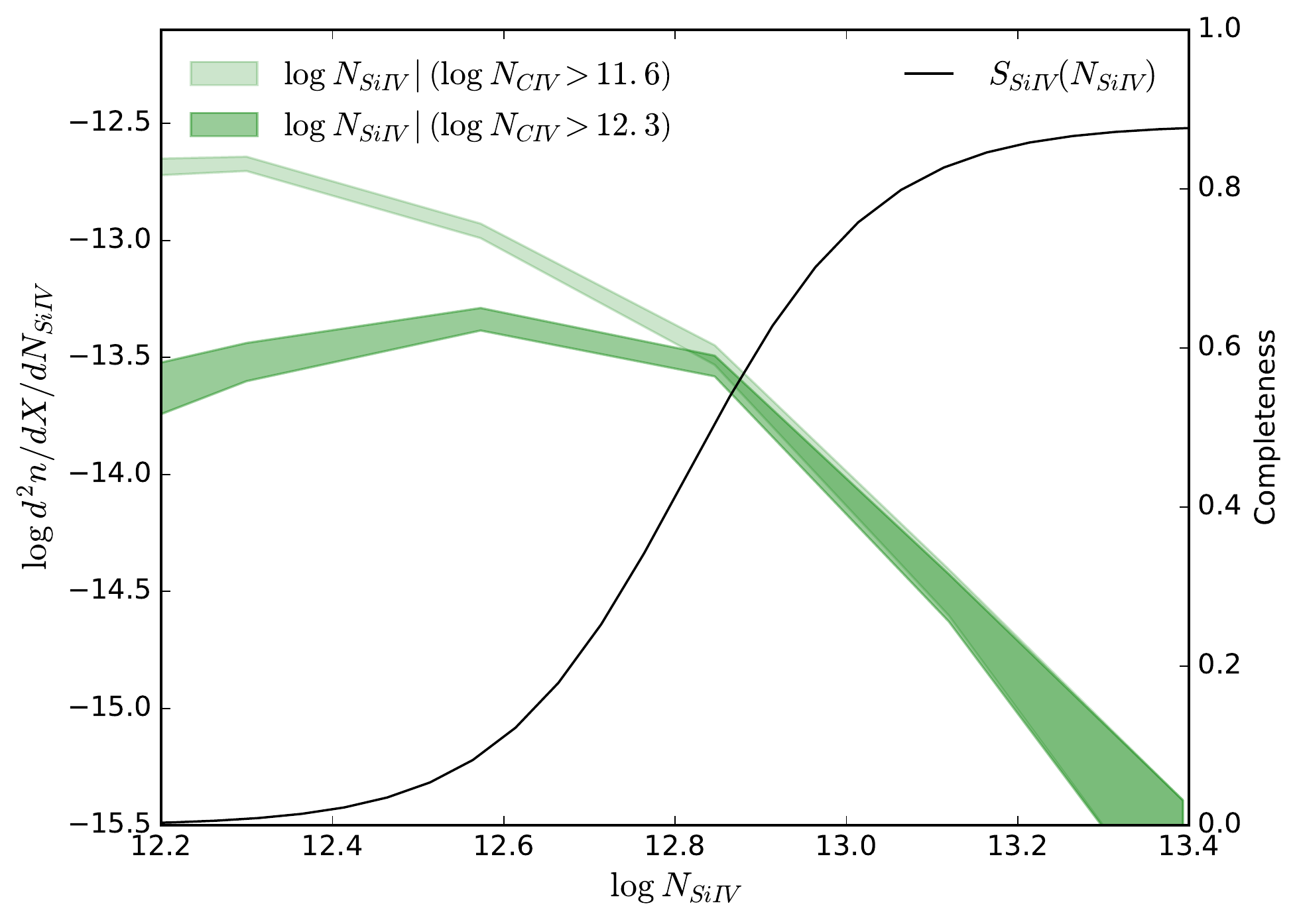}
\caption{The conditional $\siiv$ column density distribution (CCDD) for two different cutoffs in $\log N_{\civ}$, where the light green shading shows the raw distribution of $\siiv$ column densities with the cutoff set equal to the minimum column density observed in the HM12 $\civ$/$\siiv$ set of absorption systems.  The dark green shading shows a cutoff at the minimum $\log N_{\civ}$ observed in~\citet{dodo13}.  The solid black line shows the sensitivity function derived for $\siiv$ using Equation 7 from~\citet{kea16}. \label{fig:siiv_cdd}}
\end{figure}
Figure~\ref{fig:siiv_cdd} shows an example of the CCDD for $\siiv$ generated in an HM12 background given two different cutoffs in the column density of their aligned $\civ$ ($N_\mathrm{\civ, min} = 11.6$ and $12.3$) with uncertainties estimated as $\:\sqrt[]{n}$, where $n$ is the number of absorbers in the column density bin.  Within our analysis, the integral of the CCDD from $0\rightarrow\infty$ is normalized to unity; however, we plot the y-axis of Figure~\ref{fig:siiv_cdd} in absorption path length units for future reference.  We further define the sensitivity function $S_{\siiv}(N_{\siiv})$, as the probability of detecting a $\siiv$ absorber with column density $N_{\siiv}$; this takes observational limitations such as noise and line blending into account.  $f$ can be written as the integral of the CCDD weighted by the sensitivity function:
\begin{equation}\label{equation:f}
f = \int_0^\infty \frac{dP}{dN_{\siiv}} \left( N_{\siiv} \mid N_{\mathrm{\civ, min}} \right) S_{\siiv}(N_{\siiv}) \: dN_{\siiv}
\end{equation}
For example, if $S_{\mathrm{\siiv}}$ were a step function rising from 0 to 1 at $N_{\siiv} = 10^{13}$cm$^{-2}$, then $f$ would simply be the fraction of $\civ$ absorbers with $N_{\civ} > N_{\mathrm{\civ, min}}$ that also have $N_{\siiv}>10^{13}$cm$^{-2}$. 

In practice, of course, neither $dP/dN_{\siiv}$ nor $S_{\siiv}$ is known; the observed aligned fraction $f_\mathrm{data}$ is simply given as the fraction of systems that are detected in the aligned transition.  In order to measure $f_\mathrm{mod}$ in simulated catalogs, we need models for both $dP/dN_{\siiv}$ and $S_{\siiv}$.  The odds of detecting a $\siiv$ absorption system can be related to the completeness of $\civ$ detection, which is fit using the $\civ$ sensitivity function derived by~\citet{kea16}
\begin{equation}\label{eq:civ_sensitivity}
S_{\civ} \left( N_{\civ} \right) = \frac{0.881}{1 + e^{-9.1 \left(\log N_{\civ} - 13.2 \right)}}
\end{equation}
and applied to the~\citet{dodo13} data set ($5.0 \leq z \leq 6.0$).  Assuming that completeness varies with optical depth in the same way for both $\civ$ and $\siiv$, we can then assign corresponding $N_{\siiv}$ to the completeness values for various $N_{\civ}$ according to
\begin{equation}\label{equation:alignmentcut}
N_{\siiv} \equiv \frac{f_{\mathrm{\civ}} \lambda_{\civ}}{f_{\siiv} \lambda_{\siiv}}N_{\civ}
\end{equation}
where $f_{\civ}$ with $\lambda_{\civ}$ and $f_{\siiv}$ with $\lambda_{\siiv}$ are the oscillator strengths and rest frame central wavelengths of $\civ$ and $\siiv$, respectively.  This ``matched optical depth" treatment is our model for $S_{\siiv}$.

With this model for the sensitivity function, we simply approximate the aligned absorber fraction for each model $f_\mathrm{mod}$ as the fraction of simulated absorbers for which both the search and aligned transitions satisfy the optical depth-matched column density cutoffs.  This approach effectively treats the simulated catalog as a random sample from an unknown intrinsic CCDD.

\begin{figure}
\includegraphics[width=0.46 \textwidth]{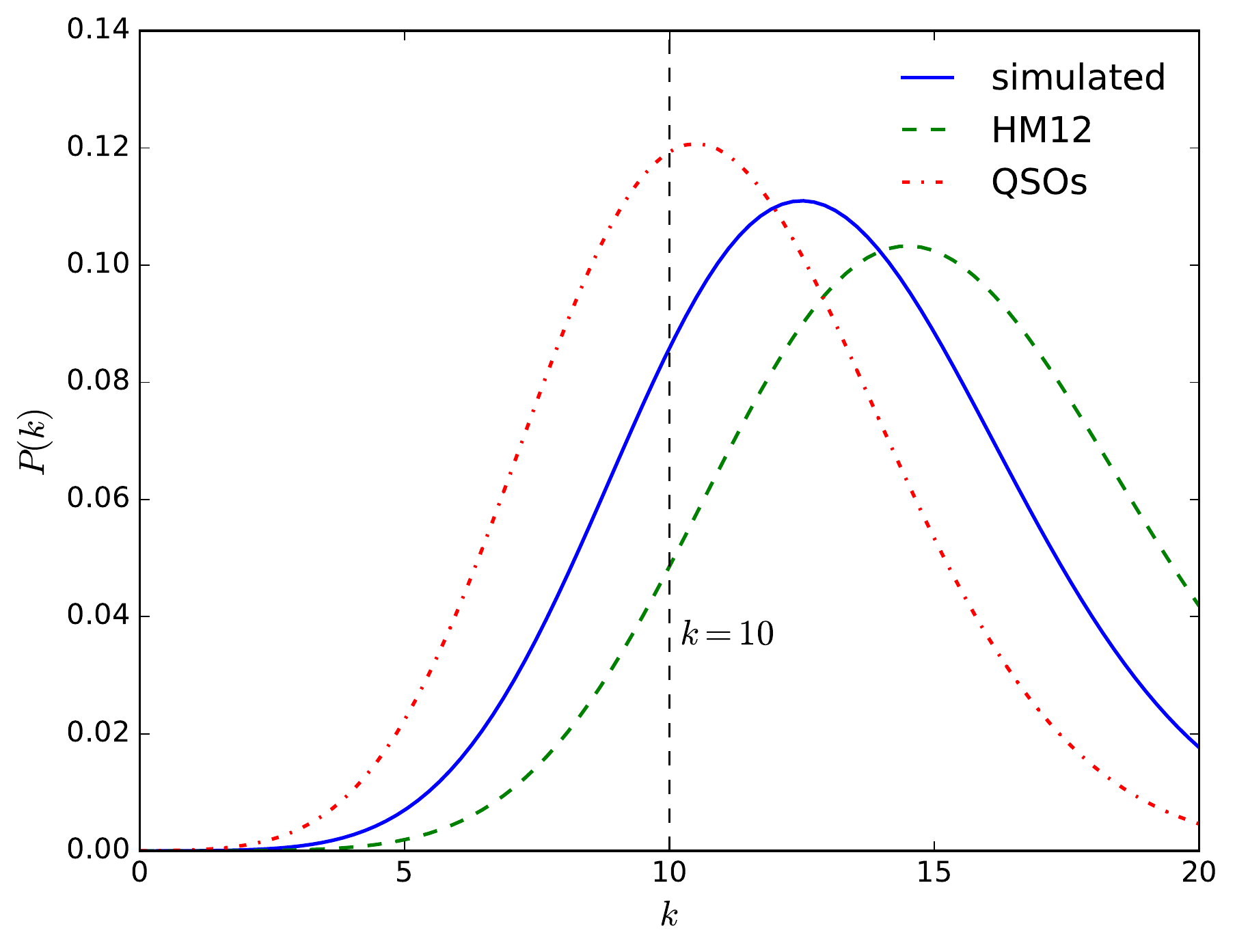}
\caption{The Poisson distribution curves for the three UVB models, where the variable $k$ is the number of observed systems satisfying the detection and aligned column density cutoffs.  The blue line shows the simulated UVB, green shows HM12, and red shows the QSOs-only UVB.  The black line emphasizes values of the curves for $\civ$/$\siiv$ systems with a $\log N_{\civ}$ cutoff of 13.25, which in the~\citet{dodo13} data results in an observed number of aligned $\siiv$ systems $k$ of 10.  For 10 observed detections, the curves shows the greatest probability $P(k)$ value corresponds to the QSOs-only UVB model. \label{fig:poisson}}
\end{figure}

Having measured $f_\mathrm{mod}$ and the equivalent value for the data, $f_\mathrm{data}$, we ask what is the probability $P$ that a survey of $\civ$ absorbers would yield the observed number of $\siiv$ absorbers.  The model that maximizes $P$ is the one favored by the data.  We assume that, for a given UVB model, $P$ is given by a Poisson distribution:
\begin{equation}\label{eqn:Poisson}
P = \frac{ \lambda^k e^{-\lambda}}{k!}
\end{equation}
Here, $\lambda$ represents the expected number of model $\civ$ systems for which an aligned $\siiv$ system can be detected; it is given by the product of $f_\mathrm{mod}$ with the total number of observed $\civ$ systems, and then rounded up to the next largest integer.  The actual number of $\civ$ systems that are also observed in the matched transition is given as $k$.  A high value of $P$ indicates that  $f_\mathrm{data}$ is very likely given the modeled population and survey characteristics.  As an example, Figure~\ref{fig:poisson} shows the Poisson distributions for aligned $\civ$/$\siiv$ systems in each UVB.  For $\log N_{\mathrm{\civ, min}} = 13.25$  with 10 observed detections, the data prefer the QSOs-only UVB.
Note that, if $f=0$---that is, only observations of upper limits are available for the aligned transition---then $k=0$ and Equation~\ref{eqn:Poisson} reduces to $P=e^{-\lambda}$.

\section{Results}\label{sec:results}
\subsection{C IV and Si IV} \label{ssec:HighIonRes}
\begin{figure*}
\centering
\includegraphics[width=0.8\textwidth]{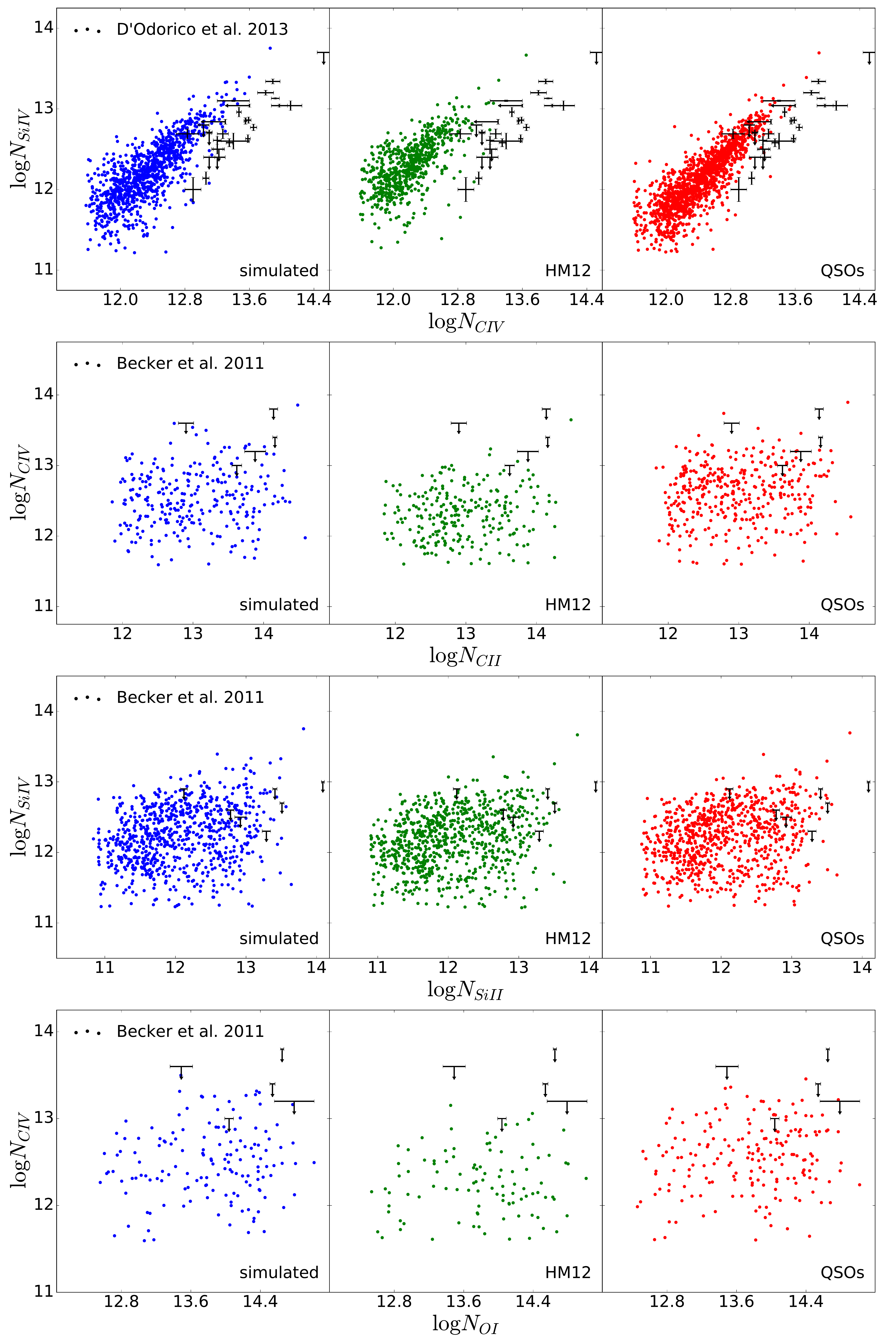}
\caption{Column densities of aligned absorber systems generated by the model UVBs (blue, green, and red for the simulated, HM12, and QSOs-only UVBs, respectively).  The black crosses show the data used for comparison for each absorber pair. \label{fig:scatterplots}}
\end{figure*}

In this section, we use the techniques presented in Section~\ref{sec:analysis} to ask whether observations can distinguish between our three UVB models.  We will begin by comparing the fraction of $\civ$-selected systems in the~\citet{dodo13} catalog that are also observed in $\siiv$.  Before delving into the statistics, however, we compare the model and observed catalogs graphically in row 1 of Figure~\ref{fig:scatterplots}.  In each panel, the colorful points show the simulated catalog for a particular UVB model while the black crosses with error bars show the observations.
There is some degree of overlap between the predicted and observed distributions at $z\sim6$ for each of the adopted UVBs.  Further, all the models recreate the slope in the $\civ$/$\siiv$ relationship quite well when compared to the data.  In detail, however, the actual column density values of the models are consistently offset from the trend given in the data, with a given column density in $\civ$ producing too much $\siiv$ (or conversely, too little $\civ$ for a given column density in $\siiv$).  This is generally true for all three UVB models.

Where the models distinguish themselves from one another is in the magnitude of the offset from the data.  Visually, it is obvious that the HM12 UVB is the least able to accurately reproduce the $\civ$/$\siiv$ values, with the bulk of its simulated absorbers falling farther from the trend than the other two models.  Comparing the results of the simulated model to the QSOs-only UVB suggests that these two UVBs are overall more similar, although there seems to be greater deviation between the predictions of the simulated UVB and the data than in the case of the QSOs-only model.

Using the method presented in Section~\ref{sec:analysis}, we compute the probability $P$ that a given model matches the observed systems.  When setting the cutoff equal to the $\civ$ column density where $S_{\civ}\left(N_{\civ} \right) = 0.5$ from Equation~\ref{eq:civ_sensitivity} (i.e., at $\log N_{\civ} = 13.25$), there is a preference for the QSOs-only UVB over the simulated UVB, at normalized probabilities of 52\% and 32\%, respectively (see Table~\ref{tab:lowionProb} for the probabilities, as well as upper and lower confidence intervals).  The HM12 UVB is disfavored, calculated at 16\%.  Even considering the distribution within 1$\mathrm{\sigma}$ of the calculated probabilities for each model, there is still a strong preference for the QSOs-only UVB, with some overlap between the models.
\begin{deluxetable}{ l c c c c }
\tabletypesize{\footnotesize}
\tablecolumns{5}
\tablewidth{\columnwidth}
\tablecaption{Normalized probabilities for all aligned pairs. The lower and upper confidence levels are given at the 16 and 84 percentiles and calculated using bootstrapping with $10^4$ iterations. \label{tab:lowionProb}}
\tablehead{
\colhead{} \vspace{-0.08cm} & \colhead{$\civ$/$\siiv$} \vspace{-0.08cm} & 
\colhead{$\cii$/$\civ$} \vspace{-0.08cm} & \colhead{$\siii$/$\siiv$} \vspace{-0.08cm} & \colhead{$\oi$/$\civ$} \vspace{-0.08cm} \\ }
\startdata
simulated & $0.32_{\:0.14}^{\:0.51}$ & $0.33_{\:0.33}^{\:0.33}$ & $0.21_{\:0.21}^{\:0.58}$ & $0.33_{\:0.33}^{\:1.00}$ \\[8pt]
HM12 & $0.16_{\:0.03}^{\:0.32}$ & $0.33_{\:0.33}^{\:0.33}$ & $0.58_{\:0.58}^{\:0.58}$ & $0.33_{\:0.33}^{\:1.00}$ \\[8pt]
QSOs-only & $0.52_{\:0.31}^{\:0.58}$ & $0.33_{\:0.33}^{\:0.33}$ & $0.21_{\:0.21}^{\:0.58}$ & $0.33_{\:0.33}^{\:0.33}$
\enddata
\tablenotetext{1}{The column density cutoff values used are as follows: $\log N_\mathrm{\civ,min} = 13.25$,  $\log N_\mathrm{\cii,min} = 12.9$, $\log N_\mathrm{\siii,min} = 12.12$, and $\log N_\mathrm{\oi,min} = 13.49$.}
\end{deluxetable}

One may wonder how the uncertainties on the carbon and silicon yields may affect the results in this section.  Observations of aligned low-ionization absorbers in~\citet{beck12} have demonstrated that the carbon to silicon abundance is consistently subsolar for $2 \leq z \leq 6$.  Treating the $\cii/\siii$ ratio as a measure of [C/Si], they report [C/Si] $= -0.19 \pm 0.10$.  We find good agreement with this value in our simulations, regardless of the UVB model.  For the HM12 background, for example, we find that observations of $\cii/\siii$ would imply [C/Si] $= -0.49 \pm 0.32$, which shows substantial overlap with the observed range.    
In detail, there may be a systematic offset to low [C/Si], which if confirmed would indicate that the simulated ratio of carbon to silicon yields is likewise low---but not by more than $\sim0.2$ dex.  Broadly, therefore, observations of $\cii/\siii$ do not encourage us to close the gap between simulated and observed $\siiv$ and $\civ$ ratios merely by adjusting the assumed metal yields.
\subsection{C II, Si II, and O I}\label{sec:LowIonRes}
Given that low-ionization absorbers trace preferentially denser gas, it is of interest to ask whether they prefer a different UVB than the purely high-ionization systems in subsection~\ref{ssec:HighIonRes}.  In order to address this question, we use the method in Section~\ref{sec:analysis} to ask whether the low-ionization systems in~\citet{beck11} can distinguish between our model UVBs, with the total number of observed systems $k$ set to 0 to accurately reflect the lack of alignment detections. 

In rows 2 through 4 of Figure~\ref{fig:scatterplots}, we show the generated model points for $\cii$/$\civ$, $\siii$/$\siiv$, and $\oi$/$\civ$, each plotted alongside the data from~\citet{beck11}.

In aligned $\cii$/$\civ$, we see that the distribution of points is broader and less linear than for $\civ$/$\siiv$, showing less of a correlation between the $\cii$ and $\civ$ column densities.  Each UVB model appears to generate a similar range of $\cii$ column densities, but the typical values of $\log N_{\civ}$ for the HM12 UVB do not extend as high as those in the other models.  The HM12 model's $\civ$ column density predictions then more typically fall below the limits defined in~\citet{beck11}, which should make its predictions more compatible with the zero reported detections.  However, this is not confirmed, as we see no statistical difference between the models (Table~\ref{tab:lowionProb}). The distinctions between the models appear to be too subtle to distinguish with so little data.

Searching for aligned $\siii$/$\siiv$, there is a wide scatter in the simulated column densities, comparable to that of $\cii$/$\civ$.  The correlations in column density between $\siii$ and $\siiv$ for all three models are somewhat more pronounced, with a higher $\siii$ column density predicting a higher $\siiv$ column density on average.  Each model $N_{\siiv}$ range overlaps the data limits, such that it is difficult to visually distinguish them.  Turning then to our statistics, we see that for aligned $\siii$/$\siiv$ there is a preference for the HM12 UVB at 58\%, while the simulated and QSOs-only UVBs are both disfavored at 21\%. However, this preference stems from a very small difference in the number of predicted systems, where two detections are predicted by both the simulated and QSOs-only models versus only a single detection by the HM12 UVB.

For aligned $\oi$/$\civ$, we see a strong reduction in the overall number of aligned systems, even compared to the other low/high ionization pairs in this section (see subsection~\ref{ssec:most_distinguishing_pairs} for more discussion of this).  There are obviously more $\civ$ column densities in violation of the limits in the case of the simulated and QSOs-only UVBs compared to HM12.  Examining the statistics, however, we once again see that there is not enough information for this method to select a preferred model, giving all three UVB models equal normalized probabilities, as well as wide ranges of likely probabilities as indicated by the confidence intervals.

Overall, it seems that the small number of observed low/high aligned transitions do not yet allow this statistical method to be used in determining a preferred UVB model for $z \sim 6$, although it will likely prove effective once more observations are available. \footnote{In this analysis, we have neglected alignments between low-ionization absorbers. In general it is expected that low-ionization absorbers are more abundant, and more likely to be aligned with one another, in locations where the local UVB is suppressed via self-shielding. Thus, such pairs would be less useful for finding the UVB characteristics coming from the sources of reionization, and more useful for examining metal comparative yields.}
\section{Discussion}\label{sec:discussion}
Our goal of the previous section was to distinguish between three trial UVBs through the detection of discrepancies between the predicted and observed column density ratio distributions, as quantified by the aligned absorber fraction and the predicted number of detections.  From the raw distributions of column densities, visual inspection would indicate that the paired high-ionization states \textit{should} prefer the QSOs-only UVB and that the low/high-ionization pairs should prefer the softer HM12.  Upon investigation of the Poisson statistics, this perception is only partially confirmed.  When we compare the observed and predicted fractions of $\civ$ absorbers with matched detections of $\siiv$, we find that $\civ$ absorbers with detected $\siiv$ prefer the QSOs-only UVB, likely due to its elevated production of $\civ$ when compared to other models.  When we compare the observed and predicted fractions of low-ionization absorbers with matched high-ionization absorbers, on the other hand, we find that $\siii$/$\siiv$ favors the HM12 model while $\cii$/$\civ$ and $\oi/\civ$ can't distinguish between any of the testing UVBs.  The preference of $\siii$/$\siiv$ absorbers for the HM12 UVB indicates that the simulated and QSOs-only UVB are overproducing the column densities of the aligned $\siiv$ compared to observations.

The disparity in the preferred UVB compared to previous work suggests that efforts to constrain the UVB from observations of metal absorbers may be sensitive to selection effects that reflect the different physical conditions in which low-, and high-ionization absorbers arise.  For example,~\citet{fin16} used the observed distributions of $\civ/\cii$ and $\civ/\siiv$ column density ratios to argue in favor of the simulated UVB.  However, their subsample of 12 systems with simultaneous constraints in three transitions was limited to strong $\civ$ absorbers, eight of which had $\siiv$ detections.  Unlike them, we do not find evidence of overproduction of $\civ$ by the QSOs-only UVB, rather consistent underproduction of $\civ$ by all three UVB models.  This result is in fair agreement with many other cosmological simulations~\citep{bir16,kea16,rahm16} that show difficulty in creating the high column densities in $\civ$ (although this trend diverges from the results of~\citet{garc17}, which showed no difficulty in reproducing 
observed $\civ$ column densities).  In contrast to our observed preference for the QSOs-only UVB, other works have demonstrated that AGN-dominated UVBs tend to overheat the IGM due to an early onset of helium reionization when compared with observations~\citep{dal17}.  In fact, it appears that the UVB's spectral slope is bracketed by simultaneous observations of the Lyman-$\alpha$ forest and of $\civ$ absorbers: The former do not permit it to grow too hard, while the latter suggest that it may harder than conventionally assumed.

Although our results warrant revisiting when larger catalogs of metal absorbers are available, we believe that even with additional observations of aligned low/high-ionization pairs, it is likely that the statistics will show variability in the preferred UVB depending on the exact ions.  In this case, the statistics of aligned metal absorbers give insight into the UVB's spatial inhomogeneity.  For example, the preference for a QSOs-only background by high-ionization pairs could indicate that the QSOs-only UVB provides a better treatment of more massive structures, which are the hosts of strong $\civ$ absorbers~\citep{kea16}.  This echoes observational evidence that bright sources may be associated with spectrally-hard UVBs~\citep{mcgr17}.

Our results confirm that the statistics of aligned absorbers are complementary to column density distributions (CDDs), where CDDs are a strong constraint on galactic outflows~\citep{opp06,bir16,kea16,rahm16}.  It has been shown that momentum-driven outflows drive enough metals into the CGM to match the observed $\civ$ CDD at $z\sim3$~\citep{opp06,bir16}.  At $z\sim6$, however, there seems to remain some tension:~\citet{kea16} examined the influences of using different simulations, wind and feedback models, and self-shielding prescriptions on metal absorber populations, showing that all combinations had difficulty in reproducing the observed $\civ$ column densities at $z\sim6$ even though $\cii$ abundances were in good agreement with observations.  They speculate that current galactic outflow prescriptions do not enrich the CGM rapidly enough, although other measures indicate that our model does not obviously show this tension~\citep{opp09,fin15,fin16}.  The better agreement of the QSOs-only and simulated UVB absorption systems with the observed $\civ$ column density at least partially reflects the tendency for the UVB to be locally enhanced near galaxies, an effect which our radiation-hydrodynamic simulations treats realistically, although~\citet{kea16} found that accounting for a stronger UVB did not boost the $\civ$ abundance by a large enough factor.

Even as our results illustrate how the statistics of aligned absorbers probe the UVB, a comparison with our previous results suggests how sensitive inferences are to the details of the input model and the analysis: Whereas~\citet{fin16} found strong evidence against both the HM12 and the QSOs-only model when considering the distribution of $\civ/\siiv$ column density ratios, we now find that high-ionization absorbers prefer the more intense QSOs-only UVB.  The difference owes to changes to our model for star formation feedback.  Broadly, the simulation used in~\citet{fin16} undersuppressed star formation and adopted a slightly stronger UVB amplitude in post-processing, leading to a more enriched and more highly-ionized CGM.  In detail, the previously-predicted star formation rate density at $z=6$ was 0.0346 $\msun$ yr$^{-1}$ Mpc$^{-3}$ (comoving) whereas our new model yields 0.0144 in the same units.  The observational compilation by~\citet{maddick14} indicates a rate of 0.01373 when converted to a~\citet{krou01} IMF, hence the newer model is likely more accurate.  Correspondingly, the volume-averaged mean metal mass fraction in Si, O, Fe, and C was $3.3\times10^{-6}$ before whereas it is only a little over half that value at $1.9\times10^{-6}$ in the new model.  Additionally, in our previous work, the simulated UVB was re-calibrated in post-processing so that its volume average matched the HM12 UVB whereas we now adopt the simulated UVB as-is.  The (re-calibrated) simulated UVB at z=6 previously corresponded to a volume-averaged HI photoionization rate of $2.52\times10^{-13}$ sec$^{-1}$ whereas our new simulation's is $1.90\times10^{-13}$ in the same units.  While both values are consistent with available observations, it is clear that adopting our new simulation's UVB will lead to less $\civ$ overall and lower $\civ/\siiv$.  The model comparison effectively compensates for this by preferring a stronger UVB than in our previous work.

A drawback to our study is that the bulk of the simulated low ionization absorbers have lower column densities than the observed systems.  This may reflect the limitations of our small simulation volume ($(12 \hmpc)^3$), which systematically underproduces the rare, strong absorbers that are most readily detected observationally.  Likewise, it is possible that our limited mass resolution leads to underproducing the dense gas reservoirs where strong low-ionization absorbers originate.  A larger dynamic range is clearly called for.  Turning this around, however, the presence of a vast population of metal absorbers that are too \textit{weak} to be observed given the detection limits of current surveys is a robust prediction.  Future surveys with next-generation telescopes and instruments, for example the GMT-Consortium Large Earth Finder on the Giant Magellan Telescope (G-CLEF on GMT)~\citep{bena16} and the High Resolution Spectrograph on the Extremely Large Telescope (HIRES on ELT)~\citep{marc16}, that probe to lower column densities will detect some of these weak absorbers, facilitating more comprehensive comparisons with the bulk of the model populations.  This could lead to a greater divergence between the different models' predictions, increasing the constraining power of observations.  In Section~\ref{sec:predictions} we investigate this possibility further.

\section{Predictions}\label{sec:predictions}
\subsection{Changes with Greater Instrument Sensitivity} \label{ssec:increasing_sensitivity}
\begin{figure*}
\centering
\includegraphics[width=\textwidth]{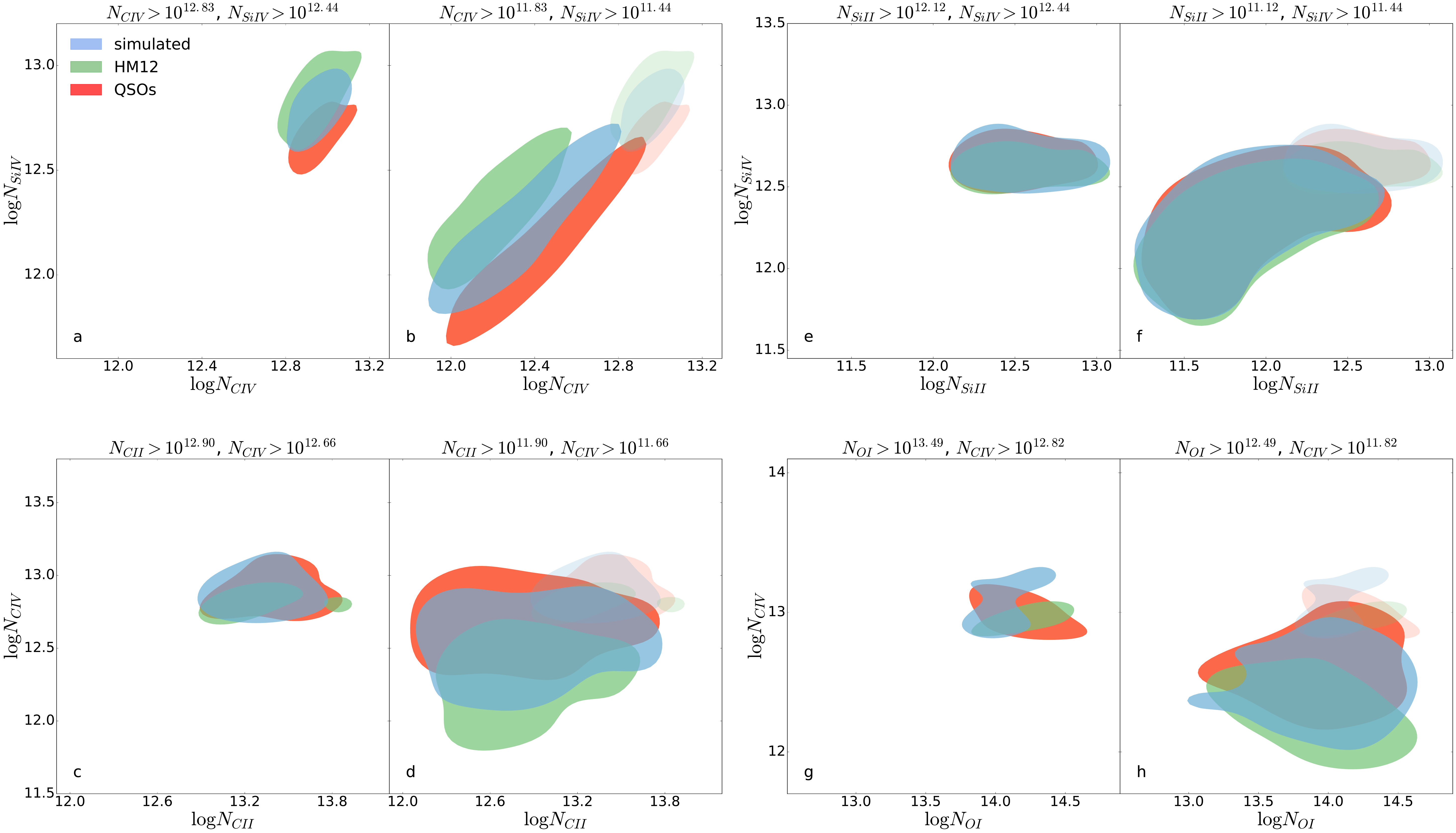}
\caption{Distributions of the aligned absorbers for each transition pair at current sensitivity levels (panels a, c, e, and g) and for a factor of ten increase in instrument sensitivity (panels b, d, f, and h), where the dark-shaded contours enclose 50 percent of the model points, as determined using a kernel density estimation.  In panels b, d, f, and h, the light-shaded contours show the distribution from the lower sensitivity limits.\label{fig:contour_predictions}}
\end{figure*}

Intuitively, one expects that weaker metal absorbers trace more diffuse gas that is located farther from the host galaxy, where the UVB is less sensitive to local fluctuations.  This picture motivates us to ask whether deeper survey detection limits would alter our ability to distinguish between our three trial UVBs.  In particular, we consider how the predicted abundance ratios change with sensitivity by defining two sensitivity levels in column density: Current sensitivity and a factor of 10 improvement.  The current sensitivity is defined as the lowest observed column density for the search ion for a given ion pair (e.g., the lowest observed column density in $\cii$ for $\cii$/$\civ$), whereas the improved sensitivity probes a factor of 10 lower in column density.  Removing all model points below that column density, contours containing 50\% of the points are then plotted for each model in Figure~\ref{fig:contour_predictions} for each defined sensitivity level.
Referring to panel \emph{a} of Figure~\ref{fig:contour_predictions}, at current sensitivity there is a small amount of overlap in the populations of aligned $\civ$ and $\siiv$ absorbers generated by the UVB models.  Looking at the distribution, there is some separation between the different models, where the harder the UVB, the greater the ratio of $N_{\civ}$ to $N_{\siiv}$.  Thus, the column density ratios for $\civ$/$\siiv$ increase as we change from an HM12, to a simulated, and to a QSOs-only UVB.  Further, the HM12 UVB shows drastically stronger $\siiv$ than the other two models.  So, even at current sensitivity levels, there is some ability to distinguish between the three UVBs based on the distribution of absorber column densities.  Incorporating an improvement in sensitivity in panel \emph{b}, we see a similar trend now continuing to lower column densities. Note that the 50\% contour boundaries become more linear, indicating that the trend likely becomes clearer as more absorbers become visible due to the sensitivity increase.  Overall, the aligned absorbers in $\civ$/$\siiv$ by the three UVBs are all quite distinguishable.

Panels \emph{c} and \emph{d} of Figure~\ref{fig:contour_predictions} show the distribution of column densities in $\cii$/$\civ$ aligned absorbers for the three UVB models.  In panel \emph{c}, which shows the distribution we would observe at current instrument sensitivity levels, there is a little variation in the typical column densities of $\cii$ and $\civ$ between the models, where $\cii$ and $\civ$ are slightly higher for the simulated and QSOs-only models.  Increasing sensitivity by a factor of 10 causes a significant change, with the representative column density in $\civ$ of the QSOs-only distribution increasing to extend above that of the other two UVBs, and the simulated model shows values which are typically higher than those in HM12.  The distributions of the QSOs-only and simulated UVBs also show higher maximum column densities in $\cii$ than the the HM12 model, and show a flatter relationship between the column densities in the two ions.  It appears that $\cii$/$\civ$ would provide greater ability to distinguish between HM12 or the simulated model versus QSOs-only if instrument sensitivity were to increase.

Examining the distribution of $\siii$/$\siiv$ absorbers for different sensitivities (panels \emph{e} and \emph{f} of Figure~\ref{fig:contour_predictions}), we see a great deal of overlap between the models.  At current sensitivity levels, there is very little variation in either the typical ranges of column densities for either ion, or in the column density ratios.  Considering a factor of 10 increase in sensitivity, the picture becomes even less heartening. There is generally an increase in the column density of $\siiv$ systems as $\siii$ column densities increase, but it is not possible to distinguish between the UVB models when considering the bulk of the points produced by the simulations.  Thus, the $\siii$/$\siiv$ pairing is not particularly sensitive to the effects of the UVB, and will not become so even with an improved ability to observe weak absorbers.

$\oi$/$\civ$ aligned absorption systems at current sensitivities (panel \emph{g} in Figure~\ref{fig:contour_predictions}) are a bit difficult to rigorously interpret because they show a peculiar distribution due to the small number of model systems satisfying the sensitivity cutoffs.  There may be an anticorrelation in the column densities of $\oi$ versus $\civ$ in the QSOs-only model, but it is difficult to be sure.  At an increased sensitivity (panel \emph{h}), there is a small offset between the HM12 and QSOs-only UVBs in $\civ$ column density, with QSOs-only showing the larger values over the entire range of $\oi$ column densities.  For both $\oi$ and $\civ$, the median column densities created by the simulated UVB are sandwiched between those of QSOs-only and HM12, generally showing  higher maxima and minima in $N_{\civ}$ than HM12.  There is now an obvious negative correlation between $N_{\civ}$ and $N_{\oi}$ in HM12 systems when observations can probe to smaller column densities.  Evidently, low-column $\oi$ absorbers trace a diffuse phase in which boosting the UVB has the effect of decreasing $N_\mathrm{\oi}$ and increasing $N_\mathrm{\civ}$ whereas high-column $\oi$ absorbers trace a self-shielded phase that is decoupled from the phase traced by $\civ$.

From these figures, we can determine:
\begin{itemize}
\item the model populations of $\civ$/$\siiv$ absorbers (Figure~\ref{fig:contour_predictions}ab) may be distinct enough that they can allow us to distinguish between UVBs even given current restrictions on instrument sensitivity;
\item aligned absorbers $\cii$/$\civ$ (Figure~\ref{fig:contour_predictions}cd) and $\oi$/$\civ$ (Figure~\ref{fig:contour_predictions}gh) could be more helpful in distinguishing between UVB traits once instrument sensitivities allow for observation at lower column densities;
\item $\siii$/$\siiv$ (Figure~\ref{fig:contour_predictions}ef) systems will not be a better probe of the UVB at greater instrument sensitivity.
\end{itemize}

It is of interest to speculate as to why the $\siii/\siiv$ ratio is so insensitive to the choice of UVB.  The ionization potentials of $\siii$ and $\siiv$ are 1.2 Ryd and 3.3 Ryd, respectively, meaning that they are both most affected by the local amplitude fluctuations that occur in a galaxy-driven UVB and are insensitive to the differences contributed by the harder spectrum from the QSOs-only UVB.  Thus, compared to the other metal absorber pairs, $\siii$/$\siiv$ is a poor probe of the hardness of the UVB.  Each ion pair including $\civ$, however, shows that the models are most distinctive in the ranges of $N_{\civ}$, so metal ions aligned with $\civ$ might be the most illuminating going forward.  We consider this further in the following section.

It is practical to state whether a factor of 10 improvement in column density is achievable with next-generation instruments on telescopes such as the ELT.  The $\civ$/$\siiv$ observations of~\citet{dodo13} were taken using the X-shooter spectrograph at the European Southern Observatory (ESO) Very Large Telescope (VLT)~\citep{vern11}.  The limiting equivalent width of an instrument is inversely proportional to the resolution $R$ times the signal-to-noise S/N per spectral resolution element.  Assuming that the limiting equivalent width is proportional to the limiting column density and leaving the S/N the same, there would need to be a factor of 10 increase in the resolution of HIRES/ELT compared to those of X-shooter/VLT.  Considering the anticipated resolution of the proposed HIRES instrument on ELT of $R\sim 100,000$, we would expect a factor of 12.5 decrease for observable $\civ$ column densities at $z\sim6$ compared to X-shooter\footnote{The improvement would be lower if considering a $\civ$ absorber at $z\leq5.5$ because the absorber would then fall in the range of the VIS arm of X-shooter, which has $R\sim11,000$, higher than that of the NIR arm}.  Currently, the lowest observed column density in $\civ$ is $\log N_{\civ} = 12.83$.  If we assume this to be the limiting column density of X-shooter, then the factor of 12.5 improvement given by HIRES/ELT would decrease the minimum observable $\civ$ column density to $\log N_{\civ} = 11.73$. For $\siiv$, the improvement would be slightly more modest, showing a factor of 9.0 decrease in observable column density of a $z\sim6$ absorber.  Again assuming a limit equal to the minimum observed aligned $\siiv$ column density $\log N_{\siiv} = 12.0$, the minimum would decrease to $\log N_{\siiv} = 11.04$ when using HIRES/ELT.  These decreases in the minimum observable $\civ$ and $\siiv$ column densities place the limits below all of our simulated systems.  Accordingly, we predict that observations of $z \sim 6$ $\civ$ and $\siiv$ should be adequate to differentiate the diverging distributions predicted by the different UVBs at lower column densities.  Observations of these aligned transitions will provide a greater ability to measure the UVB once telescopes such as the ELT are operational.

\subsection{Distinguishing Capability of Aligned Absorbers} \label{ssec:most_distinguishing_pairs}
Taking another step back, we may generalize the analysis in Section~\ref{ssec:increasing_sensitivity} by considering all possible ion combinations and asking how sensitive $f$ is to the UVB in each case.  Prior work has been done to determine what pairs of search and aligned transitions might have the most power to characterize the traits of the UVB.~\citet{opp09} considered absorption features in search ions with equivalent widths (EW) in excess of 50 $\mA$ and calculated the percentage of these systems that showed aligned ions within $\delta v \leq 30 \kms$ and $\ewlim$.  Using a similar method, though increasing $\delta v$ to 50$\kms$ to match the observational convention, we examine these percentages in Table~\ref{tab:percentages} for every possible combination of the five ions that we consider in this work, and for each of our three trial UVBs.  We identify the ion combinations showing the largest percentage differences between the models as those which are the most able to distinguish between the UVBs.

The relative numbers of aligned systems are of interest because they can tell us about the general abundance of specific ion pairs in the CGM when subject to different ionizing backgrounds.  However, the \emph{actual numbers} of systems are not insightful because they are dependent on the chosen EW strength cutoff of 50 $\mA$, and therefore on the typical column densities created in the simulation.  The small volume of this particular simulation will lead overall to smaller numbers of high column density, high EW systems when compared to larger simulations and to observations.  The percentages, however, should be invariant for the redshifts under consideration, so those values are given the most weight in this discussion.

Searching for $\oi$ systems (column 1 of Figure~\ref{tab:percentages}) demonstrates little difference between the UVBs in the numbers of aligned systems for most of the aligned ions we consider, and the numbers of $\oi$ systems are virtually identical.  Pairing $\oi$ with low-ionization absorber $\siii$ shows that roughly ninety percent of $\oi$ absorbers of $\ewlim$ will be aligned with $\siii$ regardless of the intensity or hardness of the UVB.  When aligned with the high-ionization states $\civ$ and $\siiv$, however, alignments only occur for about 0-10 percent of the $\oi$ systems.  The obvious explanation for these trends is that $\oi$ has a very low energy threshold for ionization, frequently even being used as a ``tracer'' of sorts for neutral hydrogen~\citep{oh02,kea14}.  These low ionization states are unlikely to coexist with large quantities of ionized metals, particularly in the case of the higher ionization metals in Table~\ref{tab:percentages} ($\civ$ and $\siiv$).  The only aligned ion showing significant divergence between models is $\cii$, where the percentage of alignments is identical for the simulated and QSOs-only UVBs at 78 percent, but there is a difference of nearly 20\% between these and HM12, with HM12 producing 61 percent of $\oi$ systems in alignment.

Examining detections of $\cii$ (column 2 of Figure~\ref{tab:percentages}), more absorbers are created by the QSOs-only UVB, the HM12 and simulated UVBs showing 77 percent and 93 percent of the $\cii$ absorbers seen in QSOs-only, respectively.  Each UVB generates $\cii$ alignments with $\oi$ and $\siii$ at roughly the same relative frequency as $\cii$ systems, resulting in insignificant percentage variations.  Alignments with $\civ$ and $\siiv$ occur much less frequently, being roughly twice as common with $\siiv$ as with $\civ$.  Overall, the incidence of $\cii$ and alignments with each considered transition vary similarly for each UVB, higher for the QSOs-only UVB and lower for HM12, resulting in percentages that are not distinguishable.

Searching for $\civ$ absorbers (column 3 of Figure~\ref{tab:percentages}), we see the largest discrepancy thus far between UVBs.  As the UVB model becomes harder and/or more intense the number of $\civ$ systems increases, with HM12 generating the smallest number and QSOs-only generating the largest number.  The incidence of each considered aligned absorber, with the exception of $\oi$, changes in a similar fashion.  For these systems, there are four and six times more $\civ$ absorbers generated by the simulated and QSOs-only UVB, respectively, when compared to the HM12 UVB model.  Alignments of $\civ$ with $\oi$ show the most promise, with 26 percent of simulated UVB $\civ$ systems showing alignments with $\oi$ versus 9 percent of QSOs-only systems and 0 percent of HM12 systems, distinguishing between the simulated and HM12 models.  Alignments with $\cii$ also show a significant difference between the HM12 at 14 percent and the simulated UVB at 33 percent.  Finally, $\civ$/$\siii$ shows a difference of 16 percent between the QSOs-only model at 32 percent and the simulated model at 48 percent.  These differences may be significant enough to detect observationally, thus allowing better distinction between the different models.

Searching for $\siii$ systems (column 4 of~\ref{tab:percentages}), there is virtually no variation in the number of $\siii$ absorbers between each of the models.  There are some small differences in the numbers of aligned systems with $\cii$, HM12 producing 10 percent fewer alignments than seen for QSOs-only.  Overall, this difference is likely too small to be easily detectable observationally.

Considering $\siiv$ as the search ion (column 5 of Table~\ref{tab:percentages}), HM12 UVB produces the fewest $\siiv$ systems and the simulated UVB produces the most.  For aligned $\oi$, $\cii$, and $\siii$ the number of aligned absorbers stays roughly the same, regardless of the UVB model under consideration.  However, for aligned $\siiv$/$\civ$ there is a significant difference in number of absorbers for $\civ$, particularly between the HM12 and QSOs-only model.  As a result, we see that while only 15 percent of $\siiv$ systems will have aligned $\civ$ in the presence of an HM12 UVB, 67 percent will show $\civ$ for a hard, QSOs-only-like UVB.  The simulated UVB shows less extreme results, with 40 percent of $\siiv$ systems showing aligned $\civ$.  Of all the ion transitions considered in this work, this pairing gives the strongest indicator whether the gas giving rise to the absorber pair is subject to a soft or hard UVB.  This is precisely the type of relationship that is best suited to distinguish between UVBs, namely one in which one ion changes drastically in frequency with change in the UVB (in this case $\civ$), while the other remains relatively insensitive ($\siiv$).

In summary, the analysis of these results show:
\begin{itemize}
\item most paired transitions either show no variation with the UVB, or show numbers of absorbers that vary in the same way for both transitions, meaning that there is little variation in the likelihood of alignments between models;
\item $\oi$/$\cii$, $\civ$/$\oi$, $\civ$/$\cii$, and $\civ$/$\siii$ all show variation ($\geq 15$ percent) in the percentage of aligned systems;
\item in an HM12 UVB, the probability is only 15 percent that a detected $\siiv$ system will show aligned $\civ$, significantly distinguishing it from the QSOs-only model that produces such alignments roughly 67 percent of the time.  The simulated UVB shows alignment in 40 percent of such cases, causing a large discrepancy with the QSOs-only model.
\end{itemize}
Several alignments present themselves as opportunities to help differentiate between a soft, uniform UVB and a hard, uniform UVB.  However, a much larger data set of observations at $z\sim6$ is required in order to rigorously compare the statistics of these models.

A complementary use for Table~\ref{tab:percentages} is its ability to yield estimates for the completeness of low-ionization absorber surveys at redshifts where the Lyman-$\alpha$ forest is saturated.  Observations to date~\citep{beck11,bosm17} employ a search strategy that involves looking for systems that are detected in at least two low-ionization ions simultaneously.  The incompleteness of this strategy may be estimated as the fraction of systems that are observed in one ion and unobserved in the other two.  For example, if we adopt the simulated UVB, then the fraction of OI systems that are not aligned with either Si II or CII and hence missed by this strategy is (from Table 2) $(1-0.85)\times(1-0.56)=0.066$.  Likewise, the fractions of SiII and CII systems that are undetected are 0.0168 and 0.11, respectively.  Broadly, therefore, this strategy therefore may identify $\geq90\%$ of all low-ionization ions that are stronger than the equivalent width cutoff.

\section{Conclusions}\label{sec:conclusions}
In this work, we have introduced simple statistical techniques for comparing the fraction of aligned metal absorbers in hydrodynamic simulations versus observations, and used them to explore how well current observations can constrain the slope and inhomogeneity of the UVB at $z\sim6$.  Our work builds on that of~\citet{fin16} by considering a larger suite of ions and using an improved simulation.  In contention with~\citet{fin16}, we find that the QSOs-only UVB model is preferred for the $\civ$/$\siiv$ pairing when considering all the available data.  Our main results are as follows:
\begin{itemize}
\item $\civ$/$\siiv$ absorbers show a preference for a hard, spatially-uniform UVB generated by quasars;
\item $\siii$/$\siiv$ absorbers show a tenuous preference for the HM12 UVB at $z \sim 6$;
\item $\cii$/$\civ$ and $\oi$/$\civ$ absorbers show no preference for any of the three UVB models;
\item $\siiv$/$\civ$ is the absorber pairing most sensitive to the amplitude and slope of the UVB, and should readily distinguish between the three UVBs once more observations are available.
\end{itemize}
We speculate that the apparent effect, when compared to previous work, of the choice of transition pair on the preferred UVB model reflects the tendency of different absorbers to trace different environments, with low-ionization absorbers predominantly tracing regions where the UVB is weaker while strong- and high-ionization absorbers trace regions where the UVB is stronger and/or spectrally harder.  In this view, the choice of search ion immediately imposes a bias on the inferred UVB.  It will be necessary to resolve this bias before different absorber selections can be combined to yield an overall preferred UVB, but we defer this development to future work.

In principle, our use of cosmological hydrodynamic simulations should allow us to use the overall CDDs in combination with the aligned absorber statistics to constrain simultaneously the nature of galactic outflows and of the UVB.  However, the limited dynamic range of our simulations limits our ability to leverage the strongest and most well-measured absorbers.  By further considering how aligned absorber distributions change with increasing detector sensitivity, we determined that there is extra constraining power available through an increase in detector thresholds.  It is likely that future work will rely heavily on observations from the larger telescopes that are currently under development or construction.

Overall, careful consideration is required of the effects of the ``search'' ion on the type of environment being selected, but further detailed analysis of certain aligned metal absorbers can likely distinguish between different UVBs incident to the CGM.

\section*{Acknowledgements}
Our simulations were run using \textit{Joker} at the High Performance Computing facility at New Mexico State University.  We would like to thank the ICT staff of NMSU for their invaluable technical support.  We would further like to thank Chris Churchill for his advice regarding treatment of limiting equivalent widths.  We thank George Becker, Emma Ryan-Weber, and Laura Keating for their helpful conversations.  Finally, we thank the  referee for helpful comments that improved the draft.




\bibliographystyle{mnras}
\bibliography{paper} 

\onecolumn
\begin{deluxetable}{c l r c c c c c c c c c c}
\tablecolumns{13}
\tablewidth{0pt}
\tablecaption{Aligned Absorber Percentages \label{tab:percentages}}
\tablehead{
\multicolumn{3}{c}{} & \multicolumn{10}{c}{Search Ion} \\
\colhead{} & \colhead{} & \colhead{} & \multicolumn{2}{c}{$\oi$} & \multicolumn{2}{c}{$\cii$} & \multicolumn{2}{c}{$\civ$} & \multicolumn{2}{c}{$\siii$} & \multicolumn{2}{c}{$\siiv$}\\
\colhead{} & \colhead{} & \colhead{} & $n_{\mathrm{align}}$/$n_{\mathrm{search}}$ & \% & $n_{\mathrm{align}}$/$n_{\mathrm{search}}$ & \% & $n_{\mathrm{align}}$/$n_{\mathrm{search}}$ & \% & $n_{\mathrm{align}}$/$n_{\mathrm{search}}$ & \% & $n_{\mathrm{align}}$/$n_{\mathrm{search}}$ & \%}
\startdata
\multirow{15}{*}{\rotatebox[origin=c]{90}{Aligned Ion}} & \multirow{3}{2.5em}{$\oi$} & simulated & -- & -- & 118/139 & 85 & 7/27 & 26 & 134/239 & 56 & 13/60 & 22  \\
& & HM12 & -- & -- & 91/115 & 79 & 0/7 & 0 & 136/242 & 56 & 6/33 & 18  \\
& & QSO & -- & -- & 119/149 & 80 & 4/44 & 9 & 134/239 & 56 & 7/42 & 17  \\
\cline{2-13}
& \multirow{3}{2.5em}{$\cii$} & simulated & 118/151 & 78 & -- & -- & 9/27 & 33 & 120/239 & 50 & 18/60 & 30  \\
& & HM12 & 92/152 & 61 & -- & -- & 1/7 & 14 & 103/242 & 43 & 12/33 & 36  \\
& & QSO & 119/153 & 78 & -- & -- & 11/44 & 25 & 127/239 & 53 & 17/42 & 40  \\
\cline{2-13}
& \multirow{3}{2.5em}{$\civ$} & simulated & 7/151 & 5 & 9/139 & 6 & -- & -- & 13/239 & 5 & 24/60 & 40  \\
& & HM12 & 0/152 & 0 & 1/115 & 1 & -- & -- & 3/242 & 1 & 5/33 & 15  \\
& & QSO & 6/153 & 4 & 10/149 & 7 & -- & -- & 15/239 & 6 & 28/42 & 67  \\
\cline{2-13}
& \multirow{3}{2.5em}{$\siii$} & simulated & 133/151 & 88 & 119/139 & 86 & 13/27 & 48 & -- & -- & 29/60 & 48  \\
& & HM12 & 134/152 & 88 & 103/115 & 90 & 3/7 & 43 & -- & -- & 17/33 & 52  \\
& & QSOs & 134/153 & 88 & 129/149 & 87 & 14/44 & 32 & -- & -- & 26/42 & 62  \\
\cline{2-13}
& \multirow{3}{2.5em}{$\siiv$} & simulated & 14/151 & 9 & 18/139 & 13 & 22/27 & 81 & 30/239 & 13 & -- & --  \\
& & HM12 & 6/152 & 4 & 12/115 & 10 & 5/7 & 71 & 18/242 & 7 & -- & --  \\
& & QSOs & 9/153 & 6 & 20/149 & 13 & 30/44 & 68 & 29/239 & 12 & -- & --
\enddata
\tablecomments{The fractions of absorption systems of $\ewlim$ in an aligned transition with $\delta v \leq 50 \kms$ given a detection in a search ion of $\ewlim$.}
\end{deluxetable}

\bsp	
\label{lastpage}
\end{document}